\documentclass[aps,prl,reprint,preprintnumbers,showpacs,showkeys,superscriptaddress]{revtex4-1}
\usepackage{amsmath,amssymb,bm,mathrsfs}
\usepackage{srcltx}
\usepackage{hyperref}
\usepackage{times}

\begin{document}
\title{Redundancies in Nambu--Goldstone Bosons}

\author{Haruki Watanabe}
\email{hwatanabe@berkeley.edu}
\affiliation{Department of Physics, University of California,
  Berkeley, California 94720, USA}

\author{Hitoshi Murayama}
\email{hitoshi@berkeley.edu, hitoshi.murayama@ipmu.jp}
\affiliation{Department of Physics, University of California,
  Berkeley, California 94720, USA} 
\affiliation{Theoretical Physics Group, Lawrence Berkeley National
  Laboratory, Berkeley, California 94720, USA} 
\affiliation{Kavli Institute for the Physics and Mathematics of the
  Universe (WPI), Todai Institutes for Advanced Study, University of Tokyo,
  Kashiwa 277-8583, Japan} 

\begin{abstract}
  We propose a simple criterion to identify when Nambu--Goldstone
  bosons (NGBs) for different symmetries are redundant.  It solves an
  old mystery why crystals have phonons for spontaneously broken
  translations but no gapless excitations for equally spontaneously
  broken rotations.  Similarly for a superfluid, the NGB for
  spontaneously broken Galilean symmetry is redundant with phonons.
  The most nontrivial example is Tkachenko mode for a vortex lattice
  in a superfluid, where phonons are redundant to the Tkachenko mode
  which is identified as the Boboliubov mode.
\end{abstract}

\preprint{IPMU13-0046}
\maketitle

\paragraph{Introduction.}--- In many areas of physics, it is important
to study consequences of microscopic physics on macroscopic behaviors,
sometimes called {\it emergent}\/ phenomena.  One of the best examples
in this category is the existence of gapless excitations, called
Nambu--Goldstone bosons (NGBs), when global continuous symmetries are
spontaneously broken~\cite{NambuGoldstone}.

For spontaneously broken internal symmetries in Lorentz-invariant
systems, the symmetries dictate the number ($n_{\mathrm{NGB}}$),
dispersion relation, and interactions of NGBs completely.  The present
authors have generalized this well-known results to systems without
Lorentz invariance, and proved a general
formula~\cite{WatanabeMurayama}
\begin{gather}
  n_{\mathrm{NGB}} = \mathrm{dim}(G/H) - \frac{1}{2} \mbox{rank} \rho,\label{eq:count}\\
  \rho_{ab} =-i\lim_{B \downarrow 0} \lim_{V\uparrow \infty} \frac{1}{V} \langle0| [Q_a, Q_b] |0\rangle.
  \label{eq:rho}
\end{gather}
Here, $|0\rangle$ is a ground state of the finite volume system in the presence of external fields $B_\alpha$ that favor order parameters.
Note that the symmetry
breaking pattern itself is not sufficient to fix the number of NGBs
and the additional information on the ground state, $ \rho$, is
required~\cite{footnoterho}.  Here and hereafter, whenever we refer to broken generators
$Q_a$, we mean suitable large-volume limits $\lim_{V\uparrow
  \infty} \int_V d^d x\,j_a^0(x)$, where $j^0(x)$ is the Noether charge density.

In the case of spacetime symmetries, however, the counting of NGBs is
more subtle.  Even for Lorentz-invariant systems, some examples elude the
above rule for internal symmetries, {\it e.g.}\/, spontaneously broken
conformal and scale invariance~\cite{Salam}.  There is an empirical
prescription called {\it inverse Higgs mechanism}\/ that allows one to
identify possible constraints that {\it can}\/ be imposed among
NGBs~\cite{Volkov}, while it does not dictate if they {\it should}\/
be imposed.  Little is known for theories without Lorentz invariance.

In this Letter, we propose a simple criterion to determine what
redundancies exist among NGBs in a given system.  Redundancies can arise
for two separate reasons: (1) special property of the ground state
annihilated by a linear combination of symmetry generators, and (2)
identities among Noether charge densities.  It is complementary to the
inverse Higgs mechanism because our criterion {\it requires}\/
redundancies.

This result was inspired by the work by Low and Manohar~\cite{Low},
which pointed out that a {\it local}\/ transformation of different
symmetries may lead to the same field configurations.  But they did
not clearly distinguish the classical field configurations and quantum
states and operators, and restricted themselves to Lorentz-invariant
systems.  We need to generalize their intuition and formulate it more
concretely. 

\paragraph{Noether constraints.}--- A symmetry is
spontaneously broken if its generator $Q_a$ ($a=1,\ldots,n_{\mathrm{BG}}$) has an order parameter
$\langle 0 | [Q_a, \Phi(y)] | 0 \rangle \neq 0$.  By inserting a
complete set of states, one finds the existence of a gapless state
$\langle \pi_a (\vec{p}_{a}) | j^0_a(x) | 0 \rangle \neq 0$ where
$\lim_{\vec{p}_a\rightarrow 0} E_{\pi_a}(\vec{p}_{a}) = 0$.

We first point out that the above general theorem immediately tells us
the NGBs are redundant if a linear combination of Noether
currents annihilate the ground state for non-zero coefficients $c_a$,
\begin{equation}
  \int d^dx\sum_a c_a(x) j^0_a(x) | 0\rangle = 0. \label{eq:constraints}
\end{equation}
We call them {\it Noether constraints}\/.  In general, the
coefficients $c_a(x)$ are spacetime dependent.  Since for each
spontaneously broken symmetry there must be a gapless NGB state
$|\pi_a\rangle$, let us multiply $\sum_a |\pi_a \rangle \langle \pi_a
|$ on the above equation.  Then we find that the would-be NGB states
satisfy
\begin{equation}
\int d^dx  \sum_a c_a(x) |\pi_a \rangle \langle \pi_a | j_a^0(x) | 0 \rangle =
  0. 
\end{equation}
Since $\langle \pi_a | j_a^0(x) | 0 \rangle\neq 0$ by definition, we
find $|\pi_a\rangle$ states are linearly dependent.  Therefore, the
would-be NGB states have redundancies by the number of Noether
constraints Eq.~(\ref{eq:constraints}).

The rest of the discussion is how such Noether constraints arise in
two general categories.

\paragraph{Internal Symmetries.}--- Let us first look at a simple example: Heisenberg ferromagnet.  Our argument on internal symmetry assumes the translational invariance of the ground state.  Knowing that the order parameter is the uniform magnetization, we consider the Hamiltonian $\tilde{H}=J\sum_{\langle i j\rangle}\vec{s}_i\cdot\vec{s}_j-\mu B_z\sum_i s_{zi}$ and its ground state in which all spins are alined in the positive $z$-direction.  For any $B_z>0$, the raising operator $S_+=\sum_i(s_{xi}+is_{yi})$ must annihilate the ground state, since otherwise $S_+|0\rangle$ has a lower energy ($-\mu B_z$) than the ground state.  By taking the thermodynamic limit and successively turning off the magnetic field $B_z\downarrow0$, we obtain the Noether constraint,
\begin{equation}
  \sum_i (s_{xi} + i s_{yi})| 0\rangle = 0.\label{eq:ferro}
\end{equation}
The states created by two broken charges, $S_x$ and $S_y$ are hence
not independent.  Indeed, it is known that there is only one magnon (quantized spin
wave) state, consistent with Eq.~(\ref{eq:rho}).   

In general, it is common to introduce an external field for each order parameter
so that the selected ground state has a proper thermodynamic limit.  Therefore for
antiferromagnets, we should consider the limit of vanishing staggered magnetic
field to get such a ground state.  As a result, Eq.~\eqref{eq:ferro} is not satisfied and
we can see $S_x$ and $S_y$ excite independent NGBs. 

A similar phenomenon has been discussed in a relativistic field theory with a chemical potential~\cite{MiranskySchafer}.  The model consists of a two-component scaler field $\psi(x)$ with U(2) global symmetry generated by the Pauli matrices $\tau_i$ and the identity matrix $\tau_0$.  The field acquires the vacuum expectation value, say,  $\langle0|\psi(x)|0\rangle=(v,0)^T$, breaking generators $Q_1$, $Q_2$ and $Q_3+Q_0$ spontaneously.  Due to the chemical potential, $(j^0_3+j^0_0)(x)$ also develops a non-zero expectation value.  By applying an external field to this density, we can derive the Noether constraint $\int d^dx(j_1^0+ij^0_2)(x)|0\rangle=0$, resulting in one less NGBs than the number of broken generators~\cite{MiranskySchafer}.

In order to generalize our argument to an arbitrary internal symmetry group, let us consider a simple Lie group $G$.  (Since all Lie groups can be decomposed into simple groups and U(1) factors, extension to the most general case is straightforward.)  We choose basis of generators in such a way that only $j_1^0(x)$ may have a non-vanishing expectation value~\cite{footnotej}.  Let $Q_\alpha$ ($\alpha=1,\cdots,\mathrm{rank}\,G)$ be Cartan generators of $G$ and $Q_{\pm,\sigma}=Q_{\sigma R}\pm i Q_{\sigma I}$ be raising and lowering operators such that $[Q_1, Q_{\pm,\sigma}]=\pm q_{\sigma} Q_{\pm,\sigma}$ and $q_{\sigma}>0$ ($\sigma=1,\ldots,m$).  If  $\langle0| j_1^0(x)|0\rangle$ is non-zero, it serves as an order parameter of spontaneously broken generators $Q_{\sigma R,I}$.  Hence it is legitimate to introduce an external field $B_1$ as $\tilde{H}=H-B_1 Q_1$, in addition to other external fields, if necessary.  Assuming the commutativity in taking vanishing limits for each external field, we obtain Noether constraints
\begin{equation}
\int d^dx(j^0_{\sigma R}+ij^0_{\sigma I})(x)|0\rangle=0
\end{equation}
for $\sigma=1,\ldots,m$.  As a consequence, $n_{\mathrm{NGB}}$ reduces by the number of constraints $m$.  

If we rearrange the broken generators as $Q_a=(Q_{1R},Q_{1I},\cdots,Q_{m R},Q_{m I},\cdots)$, the matrix $\rho$ defined in Eq.~\eqref{eq:rho} takes the block diagonal form where each 2 by 2 blocks reads $\lim_{B\downarrow0}\lim_{V\uparrow\infty}\langle 0|j_1^0(0)|0\rangle i\frac{q_\sigma}{2}\tau_y$. Therefore, the rank of $\rho$ is precisely $2m$, as predicted by the counting rule Eq.~\eqref{eq:count}.

\paragraph{Spacetime Symmetries.}--- Another reason for
redundancies is when the Noether charge densities are linearly
dependent.  Namely $\sum_a c_a(x) j^0_a(x)=0$ as an operator identity,
and the redundancy is obviously independent of the property of the
ground state.

To illustrate the point, let us consider a simple crystal.  The
Lagrangian or Hamiltonian is both translationally and rotationally
invariant, with six generators in three spatial dimensions.  A crystal
spontaneously breaks all six symmetries.  However, it is well-known
that there are three gapless phonon excitations (two transverse and
one longitudinal), but no more.  We are not aware of satisfactory
explanation for the lack of NGBs for rotational symmetries in the
literature.

The crucial observation is that the Noether charge densities for
translation $T^{0 i}$ and rotation $R^{0 i}$ are related by
\begin{equation}
  R^{0i} = \epsilon_{iji} x^j T^{0k}.
\end{equation}
Therefore, what could have been NGBs for spontaneously broken
rotational symmetries are redundant with those for spontaneously
broken translational symmetries, hence only three NGBs.  Note that
$x^i$ are {\it parameters}\/ and not {\it operators}\/ in quantum
field theories.  The NGB in helimagnets~\cite{Lubensky} with the Dzyaloshinskii--Moriya interaction can be understood in a similar manner.

A more nontrivial example is a superfluid.  The matter field changes
its phase under the particle-number symmetry U(1) as $\psi(\vec{x},t)
\rightarrow e^{i \theta} \psi(\vec{x},t)$, while changes both its
argument and the phase under the Galilean boost by velocity $\vec{v}$,
$\psi(\vec{x},t) \rightarrow e^{i (m \vec{v} \cdot \vec{x} -
  \frac{1}{2} m \vec{v}^2t)} \psi(\vec{x}-\vec{v} t,t)$ (we set
$\hbar=1$).  The order parameter $\langle 0 |\psi(\vec{x},t) | 0
\rangle = \psi_0$ hence breaks one phase symmetry and three boost
symmetries.  However, there is only one gapless excitation, namely the
Bogoliubov mode.  Recall that consideration of the spontaneously broken 
Galilean invariance is crucial to the Landau's criterion for superfluidity.

The lack of independent NGBs for Galilean symmetry 
again can be seen in the operator identity that the Noether
current for the Galilean boost $B^{i\mu}$ is related to the U(1)
current as
\begin{equation}
  B^{i\mu} = t T^{i\mu} - m x^i j^\mu.\label{eq:Galilean}
\end{equation}
Here and hereafter, the Greek index $\mu$ refers to the spacetime index, $0=t$, $1=x$, $2=y$, $3=z$.
It is straightforward to derive this identity from the Lagrangian density
${\cal L} = i\psi^\dagger\dot{\psi} -
    \frac{1}{2m} \nabla\psi^\dagger\nabla\psi
    - V(\psi^\dagger \psi)$.
Since the translational invariance is not broken in the superfluid,
$T^{i0}$ does not create a gapless excitation, while those created by
$B^{i0}$ and $j^0$ are linearly dependent, hence redundant.  

\paragraph{Vortex lattice.}--- 
Perhaps the most nontrivial example is the redundancy among NGBs in a vortex lattice.  
Rotating superfluids and atomic BEC form a triangular
lattice of quantized vortices~\cite{vortices}, spontaneously breaking
the translational symmetry.  It is known that the system supports a soft collective oscillation with a quadratic
dispersion, so-called Tkachenko
mode~\cite{Tkachenko,Sonin,Baym,Cornell}.  Since the Tkachenko mode is
often associated with an elliptically-polarized lattice vibration, one
may naively expect the existence of the usual (Bogoliubov) phonon,
which corresponds to the fluctuation of the superfluid phase.  Until
today, all prior works on the collective modes in the system have been
based on the hydrodynamic theory.  Although they seem to imply the
absence of such a gapless mode, the reason for the missing has been
left unclear.

To clarify the low-energy structure of the system, here we construct
an effective Lagrangian.  In order to discuss it from
the symmetry-breaking point of view, we do not take into account the
inhomogeneity due to trapping potential or the centrifugal potential.
In other words, we focus on the region where the trapping potential
almost cancels the centrifugal potential but still retains a finite
particle density.  Our system thus can be rephrased as bosons which
couple to an effective uniform magnetic field
$B_{\mathrm{eff}}=2m\Omega/e_{\mathrm{eff}}$ as if they have a charge
$e_{\mathrm{eff}}$.  The effective Lagrangian for vortices in
superfluids has been discussed in several papers~\cite{Hatsuda}, but
they did not discuss the vortex lattice configuration. They also
introduced several fields in addition to NG degrees of freedom, which
is not suitable for our purpose. 

Let us start with the standard Lagrangian~\cite{PethickSmith},
\begin{eqnarray}
\mathcal{L}&=&\frac{i}{2}(\psi^\dagger\dot{\psi}-\dot{\psi}^\dagger\psi)-\dfrac{1}{2m}|\nabla\psi|^2
\notag\\ 
&&-V_{\mathrm{trap}}(\vec{x})\psi^\dagger\psi-\frac{1}{2}g(\psi^\dagger\psi)^2. 
\end{eqnarray}
We restrict ourselves to 1+2D and the zero temperature.  To
go to the corotating frame with the angular frequency
$\vec{\Omega}=\Omega\hat{z}$, one makes the substitution
$\partial_t\rightarrow\partial_t-\vec{\Omega}\times\vec{x}\cdot\nabla$. Assuming
a Bose--Einstein condensate, we plug
$\psi=\sqrt{n}e^{-i\theta_{\mathrm{tot}}}$ into the Lagrangian and
obtain
\begin{eqnarray}
\mathcal{L}&=&\frac{i}{2}(\psi^\dagger\dot{\psi}-\dot{\psi}^\dagger\psi)-\dfrac{1}{2m}|(\nabla-im\vec{\Omega}\times\vec{x})\psi|^2\notag\\
&&-V_{\mathrm{eff}}(\vec{x})\psi^\dagger\psi
-\frac{1}{2}g(\psi^\dagger\psi)^2\notag\\
&=&n\mu-\dfrac{(\nabla n)^2}{8mn}-V_{\mathrm{eff}}(\vec{x})n-\frac{1}{2}gn^2\notag\\
&\simeq&\frac{1}{2g}[\mu-V_{\mathrm{eff}}(\vec{x})]^2, \label{eq:thirdline}
\end{eqnarray}
where $V_{\mathrm{eff}}(\vec{x})\equiv
V_{\mathrm{trap}}(\vec{x})-\dfrac{m}{2}\Omega^2x^2$ and $$\mu\equiv
\dot{\theta}_{\mathrm{tot}} -\dfrac{1}{2m}
(\nabla\theta_{\mathrm{tot}} + m\vec{\Omega}\times\vec{x})^2.$$ In the
third line (\ref{eq:thirdline}), we integrated $n$ out, keeping only
the leading term in the derivative expansion~\cite{footnote1}.

If we neglect the effective potential $V_{\mathrm{eff}}(\vec{x})$, as
we do so for the rest of the paper, the Lagrangian possesses the
\textit{magnetic} translational symmetry,
\begin{equation}
\psi'(\vec{x}+\vec{a},t)=\psi(\vec{x},t)e^{i m \vec{x}\cdot\vec{\Omega}\times\vec{a}}.
\end{equation}
Because of the lack of Galilean invariance, the energy momentum tensor
no longer satisfies $T^{0i}=m j^i$.  Instead,
\begin{equation}
T^{0i}=m j^i-2m\Omega  \epsilon^{ij} x^jj^0.\label{eq:oc}
\end{equation}
In the vortex lattice system, both $P^i\equiv \int d^d x\, T^{0i}$ and
$N\equiv \int d^d x\, j^{0}$ are spontaneously broken. However,
according to our general criterion, the operator identity
Eq.~\eqref{eq:oc} suggests that $T^{0i}$ and $j^0$ do not produce
independent NGBs.  We will explicitly verify this claim in the
following.

In the presence of vortices, the phase $\theta_{\mathrm{tot}}$
contains singularities.  We decompose $\theta_{\mathrm{tot}}$ into the
regular part $\theta_{\mathrm{reg}}$ and the vortex part
$\theta_{\mathrm{sing}}$; {\it i.e.}\/,
$\theta_{\mathrm{tot}}=\theta_{\mathrm{reg}}+\theta_{\mathrm{sing}}$.
Since $\theta_{\mathrm{sing}}$ is only defined up to a smooth
function, this decomposition is not unique and we will fix the ambiguity
later.  Due to the singularity, $\theta_{\mathrm{sing}}$ does no
longer satisfy $d^2\theta_{\mathrm{sing}}= 0$.  In fact, $*d
(d\theta_{\mathrm{sing}})$ ($*$ is the Hodge dual of $1+2$D Minkowski
space) can be identified as the vortex current $j_{\mathrm{vortex}}$
($j_{\mathrm{vortex}}^\mu=\epsilon^{\mu\nu\lambda}\partial_\nu\partial_\lambda\theta_{\mathrm{sing}}$)
that automatically satisfies the topological conservation law
$d*{j_{\mathrm{vortex}}} = \partial_\mu j_{\mathrm{vortex}}^\mu=0$.

Now let us introduce a continuum description of the vortex
dynamics. Because the crystalline order breaks the magnetic
translation, we introduce fields $X^a$ that specify the position of
the vortices. Here, we follow the notation in Ref.~\cite{Son}: $X^a$
is the Lagrangian coordinate frozen on the lattice, while $x^i$ is the
Eulerian coordinate.  We fix the relation between $X^a$ and $x^i$ in such a
way that $\vec{u}(\vec{x},t)\equiv \vec{x}-\vec{X}(\vec{x},t)$
represents the displacement from the equilibrium position $\vec{x}$.
The vortex current in the continuum description can be expressed as
$j_{\mathrm{vortex}}=*\frac{1}{2}m_0\epsilon_{ab}dX^a\wedge dX^b$
($j_{\mathrm{vortex}}^\mu=\frac{1}{2}m_0\epsilon^{\mu\nu\lambda}\epsilon_{ab}\partial_\nu
X^a\partial_\lambda X^b$)~\cite{Son} where
$\frac{m_0}{2\pi}=-\frac{m\Omega}{\pi}$~\cite{PethickSmith} is the number density
of the vortices in the equilibrium.

By equating these two expressions for the topological current, we have
$d(d\theta_{\mathrm{sing}})=-m\Omega\epsilon_{ab}dX^a\wedge dX^b$,
which gives
\begin{equation}
d\theta_{\mathrm{sing}}=-m\Omega\epsilon_{ab}X^adX^b+d\chi.
\end{equation}
A smooth function $\chi$ corresponds to the ambiguity mentioned
above. We choose $\chi=m\Omega\epsilon_{jk}x^jX^k$ so that the
explicit coordinate dependence drops from the Lagrangian. Assuming the
triangular lattice and adding the corresponding elastic energy
$E_{\mathrm{el}}(\partial\vec{u})\equiv
(2C_1+C_2)(\nabla\cdot\vec{u})^2+C_2(\nabla\times\vec{u})^2$ (in the
notation of Ref.~\cite{Baym}), we arrive at our effective Lagrangian,
\begin{eqnarray}
\mathcal{L}_{\mathrm{eff}}&=&\frac{1}{g}\mu^2-E_{\mathrm{el}}(\partial\vec{u}),\label{eq:Leff}\\
\mu&=&\dot{\theta}_{\mathrm{reg}}-m\vec{\Omega}\cdot\vec{u}\times\dot{\vec{u}}\notag\\
&&-\dfrac{1}{2m}(\nabla\theta_{\mathrm{reg}}
+2m\vec{\Omega}\times\vec{u}-m\Omega\epsilon_{kl}u^k\nabla u^l
)^2.
\end{eqnarray}
The ground state of $H-\mu_0 N$ ($N$ is the total number of particles)
is characterized as $\theta_{\mathrm{reg}}=\mu_0 t$ and $\vec{u}=0$.
$\mathcal{L}_{\mathrm{eff}}$ describes the dynamics of fluctuation
$\varphi\equiv\mu_0t-\theta_{\mathrm{reg}}$ and $\vec{u}=
\vec{x}-\vec{X}$. Similar expressions can be found in Ref.~\cite{Ikeda} that discusses the vortex lattice in superconductors in a different context, but its derivation is empirical in contrast to ours based on symmetry and derivative expansion.

As a nontrivial test, let us derive hydrodynamic equations as the
Euler-Lagrange equations of the effective Lagrangian.  Variation w.r.t
$\theta_{\mathrm{reg}}$ gives the continuity equation $\partial_\mu
j^\mu=\partial_tn+\nabla\cdot(n\vec{v})=0$, where
$n\equiv\frac{\mu}{g}$ and
\begin{eqnarray}
\vec{v}\equiv-\frac{1}{m}(\nabla\theta_{\mathrm{reg}}+2m\Omega\times\vec{u}-m\Omega\epsilon_{kl}u^k\nabla u^l).\label{eq:v}
\end{eqnarray}
Since we implicitly assumed that vortices are massless and hence
$\vec{u}$ does not have the kinetic term $\propto \dot{\vec{u}}^2$,
the Equation of Motion (EOM) of the displacement vector requires the
balance between the Magnus force and the elastic force
$\vec{F}_{\mathrm{Magnus}}+\vec{F}_{\mathrm{el}}=0$, where
$\vec{F}_{\mathrm{el}}\equiv\dfrac{\delta E_\mathrm{el}}{\delta
  \vec{u}}$ and $\vec{F}_{\mathrm{Magnus}}=
2mn\vec{\Omega}\times[\vec{v}-(\partial_t+\vec{v}\cdot\nabla)\vec{u}].$
All of these equations agree with those discussed in
Refs.~\cite{Baym,Sonin} based on the linearized hydrodynamic theory, which in
turn verifies our effective Lagrangian.  Note that our expressions are fully non-linear as required by the symmetry, {\it e.g.}\/, the third term in Eq.~(\ref{eq:v}), beyond the linearized expressions in their papers.

Let us analyze the low-energy collective mode in our effective
Lagrangian.  If we keep only quadratic terms in the fluctuation
$\varphi$ and $\vec{u}$, the Lagranigan becomes
\begin{eqnarray}
\mathcal{L}_{\mathrm{eff}}
&\simeq &\frac{n_0}{2mc_s^2}
\left[\dot{\varphi}^2-c_s^2(\partial_i\varphi+2m\Omega\epsilon_{ij}u^j)^2\right]\notag\\
&&
-n_0m\vec{\Omega}\cdot\vec{u}\times\dot{\vec{u}}
-E_{\mathrm{el}}(\partial\vec{u}).
\end{eqnarray}
In order to compare our expressions to those in the literature, we have
eliminated $g$ and $\mu_0$ in terms of the equilibrium density $n_0$
and the superfluid velocity $c_s$ by $g=\frac{\mu_0}{n_0}$ and
$\mu_0=m c_s^2$.  The remarkable feature of the effective Lagrangian
is the mass term $-2mn_0\Omega^2\vec{u}^2$.  Combined with the second
term, which makes $u^x$ and $u^y$ canonically conjugate to each other,
it explains the gapped mode with a gap $2\Omega$ in the
spectrum~\cite{Baym, Sonin}.  This mode can be identified as the collective mode with the cyclotron gap $\frac{e_{\mathrm{eff}}B_{\mathrm{eff}}}{m}=2\Omega$ predicted by Kohn's theorem~\cite{Kohn}.

Given the gap, one can safely integrate $\vec{u}$ out by using EOM, 
\begin{equation}
u^i=\frac{1}{2m\Omega}\epsilon^{ij}\partial_j\varphi+O(\partial_0\partial_i,\partial_i\partial_j\partial_k).\label{eq:latticemotion}
\end{equation}
At the leading order in the derivative expansion, the remaining Lagrangian is
\begin{eqnarray}
\mathcal{L}_{\mathrm{eff}}
\simeq \frac{n_0}{2mc_s^2}
\left[\dot{\varphi}^2-
\frac{C_2}{2mn_0}\frac{c_s^2}{\Omega^2}(\nabla^2\varphi)^2
\right],\label{eq:Tk}
\end{eqnarray}
which describes the Tkachenko mode with the dispersion relation
$E(\vec{p})=\sqrt{\frac{C_2}{2mn_0}}\frac{c_s}{\Omega}p^2+O(p^4)$~\cite{Baym,
  Sonin}.  The Tkachenko mode thus can be understood as the phase
oscillation, and the vortex lattice simply follows transverse to the motion of the
superfluid through Eq.~\eqref{eq:latticemotion}.

After all, there is only one gapless mode in the vortex lattice, as
expected from our general criterion.  In the derivation, we introduced
the redundant fields in our effective Lagrangian and observed a mass
term $\propto \vec{u}^2$ for them.  
An effective Lagrangian of crystal phonons does
not usually contain $\vec{u}$ without any derivatives, because the
invariance under the shift $\vec{u}'=\vec{u}+\vec{a}$ prohibits it. This is
why we usually expect gapless phonons~\cite{footnote2}.
However, in the current example, the appearance of the mass term does
not contradict with the symmetry --- the original magnetic translation
is still exactly realized in our effective Lagrangian
Eq.~\eqref{eq:Leff} in a nontrivial manner,
\begin{eqnarray}
\vec{u}'(\vec{x}+\vec{a},t)&=&\vec{u}(\vec{x},t)+\vec{a},\\
\theta'(\vec{x}+\vec{a},t)&=&\theta(\vec{x},t)-m\vec{a}\cdot\vec{\Omega}\times[\vec{u}(\vec{x},t)-2\vec{x}].
\end{eqnarray}
This symmetry also protects the quadratic dispersion relation of the Tkachenko mode; i.e., the lower order term
$\propto(\nabla \varphi)^2$ cannot be generated by renormalization
process in Eq.~\eqref{eq:Tk}.

It is instructive to compare the vortex lattice with a
supersolid~\cite{Son}. A supersolid exhibits a similar symmetry-breaking
pattern; namely, it breaks both (usual) translation and U(1) phase
rotation.  In contrast to the vortex lattice case, each of $d$
momentum operators $P^i$ and the number operator $N$ independently
produces a NGB, giving rise to $d+1$ NGBs in total in $d$-space
dimensions.  This is consistent with our criterion, since in the case
of supersolid, the Galilean invariance~\cite{Son} (more precisely, the
non-relativistic general-coordinate invariance~\cite{SonWingate})
leads to $T^{0i}=m j^i$.  Therefore phonons originated from the translational symmetry breaking and Bogoliubov mode are not redundant.

We appreciate fruitful discussion with T. Brauner and R. Shankar.  
We thank A. Beekman, R. Ikeda, and L. Radzihovsky for informing us Ref.~\cite{Aron},  \cite{Ikeda}, and \cite{Lubensky}, respectively.
H.W. is grateful to M. Nitta, M. Kobayashi and S. Furukawa for useful discussion on vortex lattices.
H.W. would like to appreciate the support from Honjo international scholarship
foundation.  The work of HM was supported in part by the U.S. DOE
under Contract DE-AC03-76SF00098, by the NSF under grant
PHY-1002399, the JSPS grant (C) 23540289, and by the
FIRST program Subaru Measurements of Images and
Redshifts (SuMIRe), CSTP, and by WPI, MEXT, Japan.

\paragraph{Note added.}--- 
After submitting our manuscript, the authors were informed a related preprint~ \cite{Aron}.
Although their approach is different and limited to the example of crystals, their result is consistent with our criterion.


\begin{thebibliography}{99}
\bibitem{footnoterho}
Since $\rho$ is real and antisymmetric, its rank is always even.
\bibitem{footnotej}The constant part of currents can be completely fixed by requiring the commutation relation $[Q_i,j_j^0(x)]=if_{i j}^{\,\,\,k}j_k^0(x)$.
\bibitem{footnote1}
If interested in the Bogoliubov spectrum $\omega^2=c_s^2 k^2(1+\frac{k^2}{4m^2c_s^2})$, one should not integrate $n$ out and leave it as a conjugate field to the phase.  In this case, the quantity $\mu$ in the text loses the meaning of the chemical potential.  Instead, we define the chemical potential as $\mu(n)\equiv\frac{\partial\varepsilon(n)}{\partial n}$, using the internal energy $\varepsilon(n)=\frac{1}{2g}n^2$.  The EOM of $n$ describes the time-evolution of the phase,
$\dot{\theta}_{\mathrm{reg}}=\mu(n)+m\Omega\cdot\vec{u}\times\dot{\vec{u}}+\frac{m}{2}\vec{v}^2-\frac{1}{2m\sqrt{n}}\nabla^2\sqrt{n}$.  Other EOMs remain unchanged.
\bibitem{footnote2}
In $1+2$D, the exception at the quadratic order in $\vec{u}$ is $\epsilon_{ij}u^i\dot{u}^j$, which changes by a total derivative $\partial_0(\epsilon_{ij}c^iu^j)$.
\end{thebibliography}
 \end{document}